\documentclass[twocolumn,showpacs,preprintnumbers,amsmath,amssymb,10pt,a4paper]{revtex4}

\usepackage{geometry}
\usepackage{graphicx}% Include figure files
\usepackage{dcolumn}% Align table columns on decimal point
\usepackage{bm}% bold math
\newcommand{\be}{\begin{equation}}
\newcommand{\ee}{\end{equation}}
\newcommand{\bd}{\begin{displaymath}}
\newcommand{\ed}{\end{displaymath}}
\newcommand{\BE}{\begin{eqnarray}}
\newcommand{\EE}{\end{eqnarray}}

\newcommand{\bq}{\ensuremath{\mathbf{q}}}

\newcommand{\bv}{\ensuremath{\mathbf{v}}}

\newcommand{\bx}{\ensuremath{\mathbf{x}}}

\newcommand{\bn}{\ensuremath{\mathbf{n}}}

\newcommand{\bA}{\ensuremath{\mathbf{A}}}

\newcommand{\bp}{\ensuremath{\mathbf{p}}}

\newcommand{\bchi}{{\mbox{\boldmath $\chi$}}}

\newcommand{\bkappa}{{\mbox{\boldmath $\kappa$}}}

\newcommand{\bxi}{\bm{\xi}}
\newcommand{\olx}{\overline{\mathbf{x}}}
\newcommand{\olxone}{\overline{x}_1}
\newcommand{\olxtwo}{\overline{x}_2}
\newcommand{\olxonedot}{\dot{\overline{x}}_1}
\newcommand{\olxtwodot}{\dot{\overline{x}}_2}

\newcommand{\avg}[1]{\left\langle{#1}\right\rangle}

\begin{document}

\preprint{}
\title{How limit cycles and quasi-cycles are related in systems with intrinsic noise}
% Force line breaks with \\
\author{Richard P.~Boland}
\email{richard.p.boland@postgrad.man.ac.uk}
\author{Tobias Galla}
\email{tobias.galla@manchester.ac.uk}
\author{Alan J.~McKane}
\email{alan.mckane@manchester.ac.uk}
\affiliation{Theoretical Physics Group, School of Physics and Astronomy, 
University of Manchester, Manchester M13 9PL, United Kingdom}

\date{\today}% It is always \today, today,

\begin{abstract}
Fluctuations and noise may alter the behavior of dynamical systems
considerably. For example, oscillations may be sustained by demographic 
fluctuations in biological systems where a stable fixed point is found in the 
absence of noise. We here extend the theoretical analysis of such stochastic 
effects to models which have a limit cycle for some range of the model 
parameters. We formulate a description of fluctuations about the periodic 
orbit which allows the relation between the stochastic oscillations in the
fixed point phase and the oscillations in the limit cycle phase to be 
elucidated. In the case of the limit cycle, a suitable transformation into a 
co-moving frame allow fluctuations transverse and longitudinal with respect 
to the limit cycle to be effectively decoupled. While longitudinal fluctuations
are of a diffusive nature, those in the transverse direction follow a 
stochastic path more akin to an Ornstein-Uhlenbeck process. Their power 
spectrum is computed analytically within a van Kampen expansion in the inverse 
system size. This is carried out in two different ways, and the subsequent 
comparison with numerical simulations illustrates the effects that can occur 
due to diffusion in the longitudinal direction.
\end{abstract}


\maketitle
%%%%%%%%%%%%%%%%%%%%%%%%%%%%%%%%%%%%%%%%%%%%%%%%%%%%%%%%%%%%%%%%%%%%%%%%%%%%
%%%%%%%%%%%%%%%%%%%%%%%%%%%%%%%%%%%%%%%%%%%%%%%%%%%%%%%%%%%%%%%%%%%%%%%%%%%%
\section{Introduction}

The effect of noise on nonlinear dynamical systems has been studied for some
time \cite{moss} and is now a substantial field, yet 
significant new aspects continue to be unearthed. One of the most recent of
these concerns systems which fundamentally involve discrete entities, for 
example individuals in an ecological system. These have populations which are 
modeled stochastically, for example random births and deaths. In such cases it 
may be that stochastic effects alter the behavior of non-linear systems 
substantially, and crucial differences between the properties
of a given system can be observed in the presence and in the absence of noise.
Examples can be found in the context of predator-prey population
dynamics \cite{alan1,pineda}, in evolutionary game theory \cite{frey1,frey2}, 
in cyclic trapping reactions \cite{bennaim}, in models of opinion dynamics 
\cite{lama}, in epidemics \cite{alan2,simoes,kuske} and in connection with 
genetic networks \cite{scott} or biochemical clocks \cite{gonze,alan3}. In 
these cases the ``noise'' is intrinsic to the system itself; in the parlance 
of ecology it is `demographic stochasticity', rather than environmental 
stochasticity.

One of the most intriguing effects found in these systems concerns the
existence of oscillatory behavior. It has long been conjectured that
in some situations the influence of noise due to demographic
stochasticity would be sufficient to perturb the stationary state,
predicted by a deterministic or mean-field type analysis, to produce
cyclic behavior \cite{bartlett}.  Oscillatory behaviors of this kind
are referred to as quasi-cycles \cite{nisbet}. This effect can be
demonstrated in a simple and straightforward way, and analytic results
derived show very good agreement with simulations \cite{alan1}. This
latter study conclusively demonstrates that while the deterministic
limits of models of these systems exhibit stable fixed-point behavior,
their stochastic analogues show resonant and persistent oscillations
about these fixed point solutions.

The starting point for formulating models of this kind is to define
the system with a {\em finite} number of constituents (e.g. molecules
in a chemical reaction system, individuals in the context of
population dynamics, or agents in models of social dynamics) which
interact according to a given set of possible reactions, whose
occurrence is determined by random factors (see e.g.
\cite{haken,gardiner:handbook} and references therein). In spatial
diffusive systems, for example, a certain molecular reaction may occur
only if all necessary reactants are present at a given site in space
and time, and similarly a predator in a model of population dynamics
may feed upon a unit of prey, only if both meet. Events such as birth
and death typically occur at random with Poisson statistics in such
systems, providing another source of stochasticity
\cite{gardiner:handbook}. This is demographic stochasticity. In the
limit of infinite particle numbers, such systems are faithfully
described by deterministic equations, also called rate equations. These
ordinary differential equations for concentrations of the different
reactants address the behavior of the system on a mean-field
level. Assuming a well-mixed population the rate equations are
zero-dimensional and describe uniform densities as functions of
time. The stochasticity present on the level of interactions between
individuals is averaged out in this case of an infinite system size. On
the other hand, a systematic study of first-order corrections to the
rate equations due to finite system size can be made. Such studies
have captured the behavior of fluctuations about the stationary
mean-field solution which are small enough that a linear approximation
is sufficient. In several examples, the power spectra of the
fluctuations have been computed analytically \cite{alan1,alan2,alan3}.

The aim of the present paper is to extend these existing analytical 
descriptions of finite-size stochastic effects to systems which on the
mean-field level do not always approach a stable fixed point, but instead
may tend towards a stable and periodic limit cycle solution for some range 
of the parameters. The effects of demographic stochasticity on such systems 
have been studied numerically, for example in \cite{pineda} for predator-prey 
systems with a non-linear functional response. There the shape of the resulting
auto-correlation functions of predator or prey densities have been used to help
to distinguish between noisy limit cycles and quasi-cycles. This problem could 
equivalently be analyzed by looking at the power spectra. Distinguishing 
these two types of cyclic motion might also be of relevance in the context of 
biochemical clocks \cite{gonze} and genetic networks \cite{scott}. Also of 
interest is to understand what happens at the boundary between the regimes 
where there is a stable fixed point and where there is a stable limit cycle. 
Do the cycles continuously merge with each other or are they unrelated?

To develop the analytical tools required to study such systems, we
focus on the Brusselator model 
\cite{prigogine,brown,gray,haken,serra,tomita,scott}. This model is a simple 
example of an auto-catalytic, oscillating chemical 
reaction \cite{prigogine,nicolis}. Auto-catalytic reactions are those in
which the presence of a given reactant acts to increase the rate of
its own production. A real-world realization of oscillatory chemical
reactions is given by the celebrated Belousov-Zhabotinsky reaction
\cite{zhab}. The corresponding deterministic rate equations are known to have 
limit cycle solutions provided that model parameters (i.e. reaction 
rates) are suitably chosen \cite{haken,serra}. It also has the properties 
required to show resonant oscillations in the regime where a stable fixed point
exists. Therefore it has the necessary features required for our investigation.
 
Earlier studies of the Brusselator system, e.g. by Tomita et al.
\cite{tomita} or by Scott et al. \cite{scott} have addressed finite-size 
corrections to the dynamics of this system based on van Kampen expansions of 
the corresponding master equation \cite{vankampen}, but to the best of our 
knowledge no systematic attempts have been made to study temporal correlations 
(i.e. autocorrelation functions or power spectra) of this system. 
Specifically, our objective is to build on the calculations performed in 
\cite{tomita}, and in particular to address these temporal correlations 
analytically. We also present a systematic account of different schemes of 
transformation into a co-moving frame (following the motion of the 
deterministic system around the limit cycle). We link these with different 
interpretations of stochastic simulations based on the Gillespie algorithm 
\cite{gill}. It should be noted that recent studies by Frey et al. 
\cite{frey1,frey2} also address models with limit cycles but those expansions 
are performed about unstable fixed points in the interior of the limit cycle, 
whereas an expansion about the cycle itself is carried out in the present work.

The remainder of the paper is organized along the following lines. In section
\ref{sec:bruss}, we introduce the model system initially as a mean field model 
and later as an individual based model. We then introduce the techniques of 
the van Kampen system-size expansion in Section \ref{sec:fixedpoint} by 
applying them to the case where the mean-field dynamics approach a fixed point.
A linearization about a stable limit cycle solution is carried out in 
Section \ref{sec:frenet-floquet}; we describe the limit cycle itself, the 
analysis of the Floquet multipliers and exponents and finally we discuss the 
co-moving Frenet frame. In Section \ref{sec:kampen} we use these tools to 
study the full stochastic problem of large but finite systems containing a 
limit cycle in the mean field. We derive the power spectrum describing the 
fluctuations in this case and also compare this prediction against numerical 
simulations of the individual-based Brusselator model under two different 
possible interpretations. In the final section we summarize our results and 
provide an outlook for future work.

%%%%%%%%%%%%%%%%%%%%%%%%%%%%%%%%%%%%%%%%%%%%%%%%%%%%%%%%%%%%%%%%%%%%%%%%%%%%%%
%%%%%%%%%%%%%%%%%%%%%%%%%%%%%%%%%%%%%%%%%%%%%%%%%%%%%%%%%%%%%%%%%%%%%%%%%%%%%%
\section{The Brusselator model}\label{sec:bruss}
\subsection{Deterministic mean-field rate equations}\label{subsec:bruss_mean} 
The non-spatial Brusselator model is well known in the form of the mass-action 
kinetic equations in two dynamic variables $x_1(t)\geq 0$ and $x_2(t)\geq 0$ 
representing the time-dependent concentrations of a pair of dynamic reagents. 
These equations are of the form \cite{tomita,haken,serra}
\BE\label{eq:mean}
\begin{array}{rl}
\dot{x}_1=&1-x_1(1+b-cx_1x_2),\\
\dot{x}_2=&x_1(b-cx_1x_2),
\end{array}
\EE
where $b$ and $c$ are constant and positive model parameters related to
the reaction rates. Since they will form the basis of the work we will describe
in this paper, we will briefly review the structure of these equations. Further
details may be found in textbooks on nonlinear dynamics, for instance 
\cite{serra}. For simplicity we will use the shorthand $\bx(t)=(x_1(t),x_2(t))$
for the two-dimensional vector of concentrations. 

\begin{figure}[t]\vspace{3em}
\centerline{\includegraphics[width=0.35\textwidth]{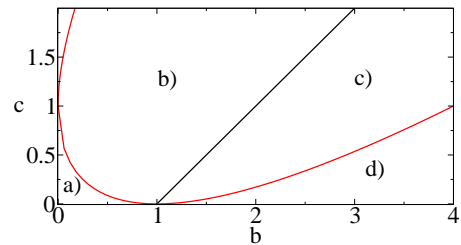}}
\caption{(Color on-line) Phase diagram of the Brusselator fixed point at 
$\mathbf{x}^{*}=(1,(b/c))$ as obtained from the deterministic 
(mean-field) theory, Eq. (\ref{eq:mean}). Behavior in the different regions 
is as follows: a) stable, non-oscillatory (both eigenvalues real and negative),
b) stable, oscillatory (eigenvalues complex, negative real part), c) unstable, 
oscillatory (eigenvalues complex, positive real part), d) unstable, 
non-oscillatory (both eigenvalues real and positive). Note that there is a 
stable limit cycle in both c) and d).}
\label{fig:pg}
\end{figure}

The two-dimensional and first-order nature of the Brusselator restricts the 
class of solutions to three possibilities. We may have fixed points at which 
the left-hand side of Eq. (\ref{eq:mean}) vanishes, limit cycles where the 
dynamics repeats in a periodic orbit or a unbounded behavior where the 
solution tends towards infinity and never returns. It turns out that this last 
possibility can be eliminated for the Brusselator if we require that the 
dynamics begins within the positive quadrant ($x_1>0$ and $x_2>0$) and also 
that we have positive and finite parameter values. A limit cycle 
does exist for a range of parameter values; this will be discussed in Section
\ref{subsec:limit-cycle}. This leaves fixed points, which we will discuss here.
It turns out that there is only one fixed point of Eq. (\ref{eq:mean}) given 
by $\bx^*=(x_1^*,x_2^*)=(1,(b/c))$. The behavior of the system close to 
this fixed point can be understood within a linear stability analysis. To this 
end, one considers a deviation 
$\varepsilon\bm{\xi}(t)\equiv\mathbf{x}(t)-\mathbf{x}^{*}$ from the fixed 
point. The pre-factor $\varepsilon$ here indicates that we assume these 
deviations to be small. Later, in the context of a van Kampen expansion, this 
parameter will take on a specific meaning in terms of the system size. Assuming
that an expansion to linear order is appropriate, one then has
\be\label{eq:fplin}
\dot\bxi=K^{*}\bxi,
\ee
where the Jacobian at the fixed point is given by
\be\label{eq:jac}
K^{*}=\left(\begin{array}{cc}
b-1 & c\\
-b & -c\end{array}\right).
\ee
The eigenvalues of the matrix $K^{*}$ determine the stability or otherwise of 
the fixed point $(x_1^*,x_2^*)$. They are both found to be real so long as 
$|b-1-c|>2\sqrt{c}$. Otherwise the eigenvalues form a complex 
conjugate pair with a real part which is negative if $b<1+c$. The line 
$b=1+c$ is a family of Hopf bifurcations which separate the parameter space 
into two phases: one in which there exists a single globally stable fixed 
point ($b<1+c$) and another in which there is a single globally stable limit 
cycle ($b>1+c$). The resulting phase diagram is depicted in Fig. \ref{fig:pg}.

%%%%%%%%%%%%%%%%%%%%%%%%%%%%%%%%%%%%%%%%%%%%%%%%%%%%%%%%%%%
\subsection{Microscopic multi-particle dynamics}\label{subsec:bruss_idv}
We can also discuss the Brusselator on the level of individual molecules. In 
this case, the system is described by four chemical reactions between four 
substances, $A, B, X_1$ and $X_2$ \cite{haken,serra,tomita}. The number of $A$ 
and $B$ particles is, by construction, constant in time, so that their 
populations do not form degrees of freedom; their role is merely to set the 
reaction rates. The state of the system at any time is therefore described by 
the number of molecules of each of the two substances $X_1$ and $X_2$, denoted 
by $n_1(t)$ and $n_2(t)$. Both of these are non-negative integers at any time 
and we will define $\bn=(n_1,n_2)$. The molecules are 
considered to interact randomly due to short-timescale fluctuations such as 
thermal activation. Hence we model their occurrence with Poisson statistics. 
As such we need only specify the expected number of occurrences per unit time 
for each reaction. We denote these transition rates as 
$T_{\nu}$ for $\nu\in\{1,\dots,4\}$, which are in turn functions of the 
population vector $\bn$. The reactions of the Brusselator are given here 
alongside their corresponding transition rate,
\begin{eqnarray}\label{eq:reactions}
\begin{array}{cccl}
A & \stackrel{}{\longrightarrow} & X_1 + A& ~~~T_{1}(\mathbf{n})=N\\
X_1 & \stackrel{}{\longrightarrow}& \emptyset & ~~~T_{2}(\mathbf{n})=n_{1}\\
X_1 + B & \stackrel{}{\longrightarrow}& X_2 + B& ~~~T_{3}(\mathbf{n})=bn_{1}\\
2X_1+X_2\phantom{.^{2}} & \stackrel{}{\longrightarrow}
& 3X_1 & ~~~T_{4}(\mathbf{n})=cN^{-2}n_{1}^{2}n_{2}.\end{array}
\end{eqnarray}
Note that different microscopic formulations of the Brusselator can be found 
in the literature (see e.g. \cite{gill,schranner,qian}), with their differences
pertaining only to the precise role of the non-dynamical substances. We here 
follow the formulation of \cite{tomita}.

The rates, $T_{\nu}(\mathbf{n})$, are derived according to the stoichiometric 
weights of each type of molecule on the left-hand side of each reaction. The 
first reaction effectively corresponds to the (spontaneous) creation of $X_1$
molecules. The rate of this reaction is proportional to the number of $A$ 
molecules which we parametrize by the integer $N$ and hence we use 
$T_1(\bn)= N$. The spontaneous decay of $X_1$ molecules, $X_1\to\emptyset$, 
occurs with a rate proportional to $n_1$ and will be set to $T_2(\bn)=n_1$. 
There is no loss of generality by doing this since we can consider this 
reaction as setting the time scale of the model. With the choices made so far, 
the first two reactions in isolation ensure that the mean number of $X_1$ 
molecules over time is $N$ (see for example \cite{gardiner:handbook}).  The 
third reaction converts molecules of type $X_1$ into type $X_2$ and occurs at 
rate $T_3(\bn)=bn_1$, where $b$ is a parameter equal to the ratio of the number
of $B$ molecules to that of $A$ molecules. The fourth reaction is of an
auto-catalytic nature (with $X_1$ being both a reactant and a product of the 
reaction) and will convert $X_2$ into $X_1$ at a rate which is quadratic in 
$n_1$ and linear in $n_2$. More precisely, the rate of this fourth reaction is 
proportional to $n_{1}(n_{1}-1)$. However, we shall assume that $n_{1}\gg1$ at 
all times and that $T_4(\bn)$ is well approximated by 
$T_4(\bn)=cN^{-2}n_1^2n_2$. The factor of $N^{-2}$ ensures that $c$ is of the 
same dimension as the parameter $b$. So long as $b$ and $c$ are independent 
of $N$, $n_2$ also scales with $N$ and hence $N$ controls the total number of 
particles in the system. This justifies the use of the label {\em system size} 
for the parameter $N$.

The stochastic time evolution of the system can be described by a master 
equation. To compactify notation we will encode the effect of an occurrence of 
reaction $\nu\in\{1,\dots,4\}$ on the system in a vector $\mathbf{v}_{\nu}$ 
describing the change of populations due to this reaction. The first 
component of $\mathbf{v}_{\nu}$ denotes the change in the number of
$X_1$ molecules due to the single occurrence of reaction $\nu$, and the second 
component the change in the number of $X_2$ molecules. For example, an 
occurrence of the first reaction increases $n_{1}$ by one while leaving 
$n_{2}$ unchanged. In all we have,
\BE
\mathbf{v}_{1}=(1,0),&~~&\mathbf{v}_{2}=(-1,0),\nonumber\\
\mathbf{v}_{3}=(-1,1),&~~~&\mathbf{v}_{4}=(1,-1).\label{eq:changevectors}
\EE 
The evolution of the time-dependent probability,
$P_{\bn}(t)$, of finding the system in state $\bn=(n_1,n_2)$ at time
$t$ is described by the master equation,
\begin{equation}
\frac{\mathrm{d}}{\mathrm{d}t}P_{\mathbf{n}}(t)=
\sum_{\nu=1}^4\left(T_{\nu}(\mathbf{n}-\mathbf{v}_{\nu})
P_{\mathbf{n}-\mathbf{v}_{\nu}}(t)-T_{\nu}(\mathbf{n})
P_{\mathbf{n}}(t)\right)\label{eq:Masterequation},
\end{equation}
subject to initial conditions $P_{\bf n}(t_0)$. The first term in the 
summation accounts for transitions of the system state from
$\mathbf{n}-\mathbf{v}_\nu\rightarrow \mathbf{n}$ while the second term
accounts for transitions away from the state $\mathbf{n}$.

While the problem of solving the master equation exactly is intractable, the 
mean-field behavior of the system, Eq. (\ref{eq:mean}), may be recovered by 
multiplying the master equation on both sides by the population vector $\bn$ 
followed by a summation over all possible configuration states. This leads to
\be
\frac{\mathrm{d}}{\mathrm{d}t}{\left\langle \mathbf{n}\right\rangle }=
\sum_{\nu=1}^4\mathbf{v}_{\nu}\left\langle T_{\nu}(\mathbf{n})\right\rangle 
\label{eq:meanfield},
\ee
where the brackets $\avg{\dots}$ denote a time-dependent ensemble average over 
realizations of the stochastic dynamics, i.e. 
$\avg{f(\mathbf{n})}=\sum_{\mathbf{n}}P_{\mathbf{n}}(t)f(\mathbf{n})$ for any 
function $f(\mathbf{n})$ of the state-vector. We shall now also use the 
mean-field approximation $\avg{T_{\nu}(\bn)}\approx T_{\nu}(\avg{\bn})$, which 
amounts to neglecting correlations by replacing $\avg{n_1^2 n_2}$ with 
$\avg{n_1}^2\avg{n_2}$. To simplify notation further we define the 
dimensionless variable $\mathbf{x}(t)\equiv N^{-1}\avg{\mathbf{n}(t)}$ and 
dimensionless transition rates $a_{\nu}(\bx)$ through the identity 
$Na_{\nu}(\bx)\equiv T_{\nu}(N\bx)$ for all reactions. Finally, the rate of 
change of the mean-field concentration is
\be
\dot{\bx}=\mathbf{A}(\bx)=\sum_\nu\bv_\nu a_{\nu}(\bx),
\ee
where $\nu$ runs over all reactions. In the Brusselator we have
\BE
a_{1}(\bx)=1,&~~&a_{2}(\bx)=x_1,\nonumber\\
a_{3}(\bx)=bx_1,&~~~&a_{4}(\bx)=cx_1^2x_2,\label{eq:dimlessrates}
\EE
and so we have the functions $A_1(\mathbf{x})=1-x_1(1+b-cx_1x_2)$ and 
$A_2(\mathbf{x})=x_1(b-cx_1x_2)$ which agrees with the right-hand sides of the 
mass-action equations (see Eq. (\ref{eq:mean})). We shall see that the 
definition of $a_{\nu}(\bx)$ finds further use in the following section.
%%%%%%%%%%%%%%%%%%%%%%%%%%%%%%%%%%%%%%%%%%%%%%%%%%%%%%%%%%%%%%%%%%%%
\section{Stochastic effects in the fixed-point phase}\label{sec:fixedpoint}
One of the aims of this paper is to understand how the cycles generated by
stochastic amplification, in the regime where a stable fixed point exists,
behave as parameters change so that the fixed point becomes unstable and a 
limit cycle is born. Therefore in this section we examine the nature of these 
stochastic cycles by restricting the analysis to choices of the model 
parameters corresponding to points in phase b) of Fig. \ref{fig:pg}. 

In the previous section we showed that the average dynamics of a large
ensemble of the finite Brusselator system, defined by the reaction
dynamics of Eq. (\ref{eq:reactions}), will follow the deterministic
path laid out by Eq. (\ref{eq:mean}). Individual realizations will, of
course, follow random paths but simulations show these will move
towards and subsequently stay close to the fixed point. In particular
the random fluctuations away from a stable fixed point solution are of
relative order $1/\sqrt{N}$, as has been observed in similar analysis
of other reaction systems \cite{alan1,alan2,alan3,tomita}.  The
characteristics of the fluctuations can be studied analytically by
means of an expansion of the master equation in the inverse system
size \cite{vankampen}. This is a standard tool in the analysis of
interacting particle systems, commonly referred to as van Kampen's
system-size expansion, which has been applied to a number of different
systems e.g. in \cite{alan1,alan2,alan3,tomita}, so we will not present
the full details of the mathematical analysis here. Some of the intermediate
steps are reported in the Appendix. The key idea is to write the particle 
populations $n_1$ and $n_2$ of the finite system as 
\BE
n_1/N&=&x_1^*+\xi_1/\sqrt{N},\nonumber\\
n_2/N&=&x_2^*+\xi_2/\sqrt{N}.
\label{eq:SSE_FixedPoint} 
\EE 
It follows that $\xi_1$ and $\xi_2$ are also random variables which
represent the fluctuations of the dynamics of the finite system about
the stationary solution of the mean-field equations. We will
frequently use the shorthand $\bm{\xi}=(\xi_1,\xi_2)$. A
systematic expansion of the master equation (\ref{eq:Masterequation})
in powers of $N^{-1/2}$ is then carried out following the lines of
\cite{vankampen}. The leading order terms yield the deterministic
mean-field equations while the next-to-leading order terms give rise
to a linear Fokker-Planck equation (see Eq. (\ref{eq:FP})), describing
the time evolution of the probability density function of
$\bm{\xi}$. The corresponding drift matrix is the Jacobian
$K^*$ of the deterministic dynamics (as defined in
Eq. (\ref{eq:jac})). The appearance of $K^*$ should not be surprising:
Eq. (\ref{eq:SSE_FixedPoint}) has the same form used in the linear
stability analysis of Section \ref{subsec:bruss_mean}, with $\epsilon
= 1/\sqrt{N}$. The diffusion matrix, $D^*$, in the Fokker-Planck equation is 
given by Eq. (\ref{eq:Dform}) evaluated at the fixed point. This yields, 
\be 
D^*=\left(\begin{array}{cc}(1+b)& -b\\ -b & b
\end{array}\right).  
\ee 
The Fokker-Planck equation (\ref{eq:FP}) is
equivalent to a Langevin system of the form 
\be
\label{eq:langevinfp}
\dot{\bm{\xi}}(t)=K^*\bm{\xi}(t)+\mathbf{f}(t), 
\ee 
where $\mathbf{f}(t)=(f_1(t),f_2(t))$ is bivariate Gaussian white noise of
zero mean and with the following co-variance matrix indicating
correlations between components: 
\be 
\left\langle f_i(t)f_j(t^\prime)\right\rangle= 
2D_{ij}^*\delta(t-t^\prime) ~~~~~~~~ i,j\in\{1,2\}.  
\ee 

\begin{figure}[t]
\centerline{\includegraphics[bb=-90bp 30bp 528bp 430bp,clip,width=1\columnwidth]{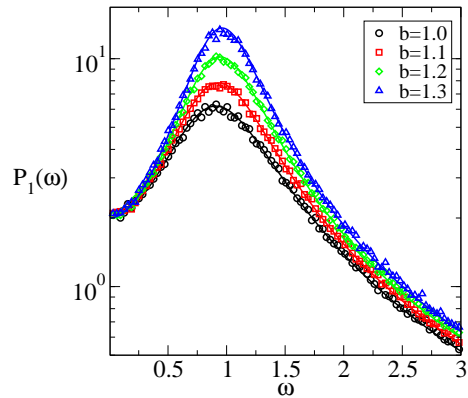}}
\caption{(Color on-line) Power spectrum $P_1(\omega)=\avg{|\widetilde{\xi_1}
(\omega)|^2}$ of fluctuations in the concentrations of $X_1$ molecules in the 
fixed-point phase ($c=1,b=1.8,1.85,1.9,1.95$ from bottom to top at the 
maximum). Solid lines show results from the analytical theory, markers are 
from stochastic simulations using Gillespie's algorithm (simulations are run 
up to $t_\mathrm{f}=150$, system size is $N=10^5$, averages over $10^4$ 
samples are taken).}
\label{fig:fluctfp}
\end{figure}

Due to the linear character of
Eq. (\ref{eq:langevinfp}) and given that the drift matrix is constant
in time it is straightforward to obtain analytical expressions for the
power spectra $P_1(\omega)=\avg{|\widetilde{\xi_1}(\omega)|^2}$ and
$P_2(\omega)=\avg{|\widetilde{\xi_2}(\omega)|^2}$. We have here
written $\widetilde{\xi_i}(\omega)$ for the Fourier transform of the
fluctuating variables $\xi_i(t)$ ($i=1,2$) with, \be
\widetilde{\xi_i}(\omega)=\int_{-\infty}^\infty \xi_i(t)
e^{-\mathrm{i}\omega t}\mathrm{d}t.  \ee Following the steps of
\cite{alan1,alan2,alan3} one finds 
\BE
P_1(\omega)=&2\left((1+b)\omega^2+c^2\right)\mathcal{D}^{-1}(\omega),\\
P_2(\omega)=&2b\left(\omega^2+1+b\right)\mathcal{D}^{-1}(\omega),\\
\mathcal{D}(\omega)=&{(c-\omega^2)^2+(1+c-b)^2\omega^2} 
\label{denom}.
\EE 
As seen in Fig. \ref{fig:fluctfp} these spectra each show a maximum at a 
non-zero frequency, indicating amplified coherent oscillations due to the 
demographic noise. These analytical predictions compare well against 
simulations for different values of the model parameters $b$ and $c$ well 
inside the fixed-point phase. Numerical estimates for the power spectra are 
obtained through the repeated simulation of the microscopic chemical reactions 
using the Gillespie algorithm \cite{gill}. This is a widely used method to
sample random paths from the solution to a master equation derived for 
Markovian particle systems. Only as the boundary of the fixed point phase is 
approached (i.e. as $b\to 2$ from below for the fixed value of $c=1$) do 
systematic deviations between the theory and Gillespie simulations emerge 
visibly in Fig.
\begin{figure}[t]
\centerline{\includegraphics[bb=-90bp 30bp 528bp 430bp,clip,width=1\columnwidth]{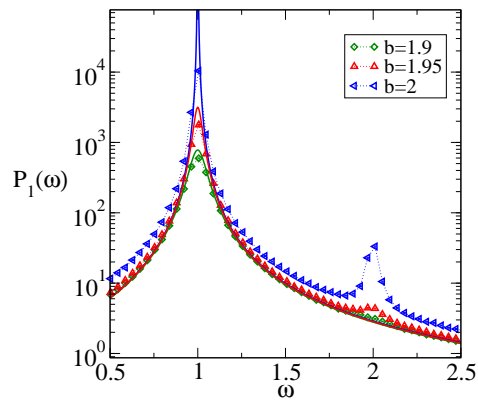}}
\caption{(Color on-line) Power spectrum $P_1(\omega)=\avg{|\widetilde{\xi_1}
(\omega)|^2}$ of fluctuations in the concentrations of $X_1$ molecules in the 
fixed-point phase near the onset of limit cycle behavior. Solid lines show 
results from the analytical theory, markers are from stochastic simulations 
(run up to $t_\mathrm{f}=150$, system size is $N=10^5$, averages over 
$10^4$ samples are taken).}
\label{fig:fluctfp2}
\end{figure}
\ref{fig:fluctfp2}. In particular, the power spectra from simulations 
begin to exhibit peaks at harmonics of the fundamental frequency given by the 
first maximum, which are not captured by the linear theory. These are the 
early precursors to the onset of limit cycles due to the stochastic broadening 
of the Hopf bifurcation. Further discussion of the effect of stochasticity on 
a Hopf bifurcation can be found in \cite{arnold}. 

One of the main points of interest is to see what happens as the boundary of
the fixed point phase is reached. When $b=1.95$, for example, the peak can be 
seen to reach a height of $~3\times 10^3$. In fact, it appears from 
Fig. \ref{fig:fluctfp2} that as $b\to 2$ the maximum of $P_1(\omega)$ tends 
towards infinity. Stochastic effects and resonant amplification of oscillations
can hence become macroscopic (i.e. of the same order of magnitude as the 
mean-field dynamics) for systems of very large system size, close to the 
transition into limit cycle behavior. This is an extreme case of stochastic 
amplification due to a resonance as can be seen from Eq. (\ref{denom}): if $c$ 
is set equal to 1 and $b=2-\delta$, the denominator vanishes at frequencies 
given by $\omega^2 = [1 - (\delta^{2}/2)] \pm i\delta$. When $b<2$, there
is no zero for real $\omega$, however as $b \to 2$, the pole approaches the 
real axis and becomes real at $\omega =1$ when $b=2$. The linear stability
analysis can be extended into region $b>2$: for $b=2+\tilde{\delta}$,
there is an unstable spiral of period $2\pi$. For very small $\tilde{\delta}$,
when the exponential growth can be neglected, this is a center of period 
$2\pi$, which is the nascent limit cycle --- not to be confused with a 
perturbation about the limit cycle to be discussed shortly.

The next section will discuss general technical details of how to characterize 
the stability of limit cycles in dynamical systems, and we will in particular 
review elements of Floquet theory and Frenet co-moving frames. Both 
of these are standard tools used to study dynamical systems exhibiting limit 
cycles, and are as such not directly concerned with stochastic effects, but 
with perturbations and fluctuations about periodic attractors in general. We 
will return to stochastic systems in Section \ref{sec:kampen}, before 
conclusions will be drawn in Section \ref{sec:discussion}. 
%%%%%%%%%%%%%%%%%%%%%%%%%%%%%%%%%%%%%%%%%%%%%%%%%%%%%%%%%%%%%%%%%%%%%%%%%%%%%%
%%%%%%%%%%%%%%%%%%%%%%%%%%%%%%%%%%%%%%%%%%%%%%%%%%%%%%%%%%%%%%%%%%%%%%%%%%%%%%
\section{Floquet theory and rotation into Frenet frame}
\label{sec:frenet-floquet}

%%%%%%%%%%%%%%%%%%%%%%%%%%%%%%%%%%%%%%%%%%%%%%%%%%%%%%%%%%%%%%%%%%%%%%%%%%%%%%
\subsection{Limit cycles in the Brusselator system}\label{subsec:limit-cycle}
\begin{figure}
\includegraphics[clip,width=.7\columnwidth,keepaspectratio]{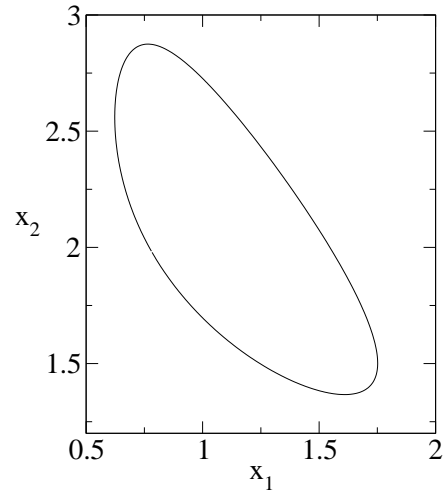}
\caption{\label{fig:cycle} (Color on-line) Illustration of the limit cycle 
solution ($\olxone(t),\olxtwo(t)$) of the deterministic Brusselator system at
fixed parameters $b=2.2, c=1$.}
\end{figure}

In the phases labeled by c) and d) in the phase diagram
(Fig. \ref{fig:pg}), the deterministic Brusselator system, as
described by Eq. (\ref{eq:mean}), exhibits limit-cycle behavior since
we can eliminate both stable fixed points and unbounded trajectories. For
positive initial conditions, Eq. (\ref{eq:mean}) admits a stable periodic
solution of period $T$. The period $T$ will generally depend on the
choice of model parameters $b$ and $c$. It is found that, although $T$
changes significantly with $c$, if we set $c=1$ as we do in this
paper, and change $b$ only in a narrow band about $b=2$, the period
will change little from the value of $2\pi$ found in the last section.
Thus in what follows we will find that the angular frequency of the
limit cycle remains close to $\omega=1$.
 
We will label limit cycle solutions by
$\mathbf{\overline{x}}(t)=(\overline{x}_1(t),\overline{x}_2(t))$ in
the following calculations and have 
$\mathbf{\overline{x}}(t+T)=\mathbf{\overline{x}}(t)$ for all times, $t$. In
general the curve, $\mathbf{\overline{x}}(t)$, cannot be calculated in closed 
form.  However good estimates can be obtained via numerical integration of 
Eq. (\ref{eq:mean}). The geometrical shape of the limit cycle of the 
Brusselator model is illustrated for a fixed choice of the model parameters 
$b$ and $c$ in Fig. \ref{fig:cycle}.

In order to study stability, we now consider a dynamical path beginning close 
to, but not on, the limit cycle, $\olx(t)$. If the limit cycle solution is 
stable then the difference between this path and the geometric curve of the 
limit cycle will decay as time progresses. Similarly to the expansion about a 
fixed point, we can write this difference as
\be
\varepsilon\bm{\xi}(t)=\mathbf{x}(t)-\mathbf{\overline{x}}(t)
\label{eq:displacement}
\ee
where, again, $\varepsilon$ expresses our anticipation that the deviation 
from the limit cycle is small. Expanding Eq. (\ref{eq:mean}) in powers of
$\varepsilon$ and letting $\varepsilon\to 0$, one then finds that the 
time evolution of $\bxi(t)$ takes on the linear form, 
\begin{equation}
\frac{\mathrm{d}}{\mathrm{d}t}\bxi(t)=K(t)\bxi(t),\label{eq:floquet}
\end{equation}
where the matrix $K(t)$ is found to take the specific form of 
Eq. (\ref{eq:Kform}) for the Brusselator model. Studying the local stability 
of limit cycle solutions against perturbation is hence the analogue of 
studying the stability of fixed points as discussed above. The elements of 
$K(t)$ are given by $K_{ij}(t)=K_{ij}(\olx(t))$ which is simply the 
matrix, $K_{ij}(\mathbf{x})=\partial a_{\nu}(\mathbf{x})/\partial x_j$ 
evaluated at the limit cycle. Therefore, due to the periodic nature of 
$\olx(t)$, all elements of $K(t)$ are periodic. 
%%%%%%%%%%%%%%%%%%%%%%%%%%%%%%%%%%%%%%%%%%%%%%%%%%%%%%%%%%%%%%%%%%%%%%%%%%%%%%
\subsection{Floquet Theory}\label{subsec:floq}
An analytical tool to characterize the stability or otherwise of limit
cycle solutions is Floquet theory --- the mathematical theory of linear 
differential equations with periodic coefficients (see \cite{grimshaw}, whose 
notation we will use). Since, in Eq. (\ref{eq:floquet}), we have 
$K(t+T)=K(t)$, Floquet theory is applicable. In our case, $T$ is the period of 
the mean-field limit cycle under consideration. 

In essence Floquet theory states that, provided $X(t)$ is a
fundamental matrix of the system (\ref{eq:floquet}), then there exists a
non-singular constant matrix $B$ such that \be X(t+T)=X(t)B \ee for
all $t$. In addition one has 
\be
\label{eq:det} 
\det B=\exp\left(\int_0^T \mbox{tr} K(t)\mathrm{d}t\right).  
\ee 
While the matrix $B$ in general depends on the choice of the particular
fundamental matrix $X(t)$, its eigenvalues (and determinant) do
not. The eigenvalues of $B$ are usually referred to as the Floquet
multipliers of the system (\ref{eq:floquet}). In the case of the
Brusselator the matrices $K(t),X(t)$ and $B$ are $2\times 2$ matrices
and we denote the resulting Floquet multipliers by $\rho_1$ and
$\rho_2$. Characteristic exponents $\mu_1$ and $\mu_2$ are then
defined by setting $\rho_i=e^{\mu_i T}$ for $i\in\{1,2\}$. Further
results of Floquet theory then concern the solutions of
(\ref{eq:floquet}). If $\rho$ is a characteristic multiplier for
(\ref{eq:floquet}) and $\mu$ the corresponding exponent then it can
be shown that there exists a particular solution $\bxi(t)$ of
(\ref{eq:floquet}), such that 
\be 
\bxi(t+T)=\rho\bxi(t)~~\forall t.
\ee 
One then finds that this solution can be expressed in terms of a periodic 
function $\bp(t)$ (i.e. one with $\bp(t+T)=\bp(t)$) scaled by an exponential, 
\be 
\bxi(t)=e^{\mu t} \bp(t).  
\ee 
General solutions of (\ref{eq:floquet}) can therefore be written as a linear
combination of functions of this form. For example in our
two-dimensional system, 
\be 
\bxi(t)=c_1 e^{\mu_1 t} \bp^{(1)}(t)+c_2 e^{\mu_2 t} \bp^{(2)}(t), 
\ee 
with $c_1,c_2$ constant coefficients determined by initial conditions.

\begin{figure}
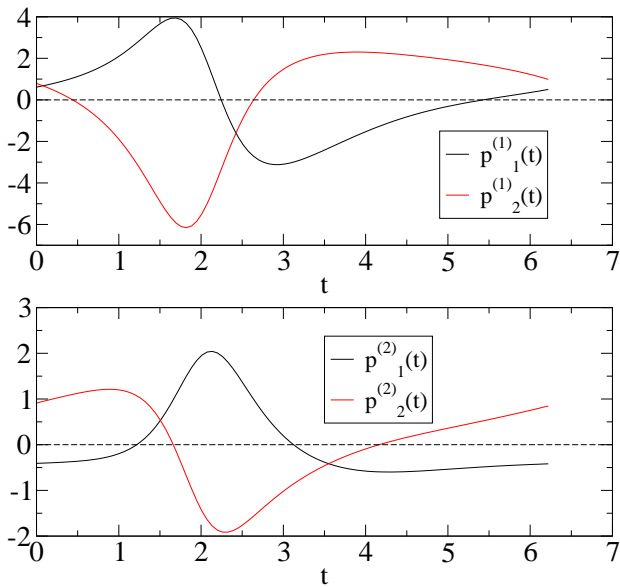

\includegraphics[clip,width=1\columnwidth,keepaspectratio]
{fig5a.eps}
\includegraphics[clip,width=1\columnwidth,keepaspectratio]
{fig5b.eps}
\caption{\label{fig:floquetCart}(Color on-line) Periodic functions of the 
Floquet analysis of Eq. (\ref{eq:floquet}). The curves show each component 
of the vector-valued functions $\bp^{(1)}(t)$ and $\bp^{(2)}(t)$ respectively. 
The function associated with the vanishing exponent is displayed in the upper 
graph, while that associated with the negative real exponent is displayed in
the lower graph.}
\end{figure}

The Floquet analysis simplifies for the class of problems where the
linear differential equations (\ref{eq:floquet}) are derived from a
dynamical system, $\dot{\bx}(t)=\bA(\bx)$, with a limit cycle
$\olx(t)$. In this case, it is easy to see by differentiation of the
original equation of motion that the vector of velocities,
$\dot\olx(t)=(\dot{\overline{x}}_1(t),\dot{\overline{x}}_2(t))$, is a
solution to (\ref{eq:floquet}). Since the velocity vector itself is a
periodic function of time, we are therefore assured that one of
the Floquet multipliers is equal to unity, $\rho_1=1$.  That is, the
corresponding exponent, $\mu_1$, vanishes. This is a general result
for all linear expansions about limit cycles arising from first-order
equations. The remaining eigenvalue of $B$ can then be determined
using Eq. (\ref{eq:det}) and specifically for the Brusselator system
we find that the corresponding Floquet exponent is given by 
\be
\mu_2=\frac{1}{T}\int_0^T (-1-b+2c\olxone(t)\olxtwo(t)-c\olxone(t)^2)
~ \mathrm{d}t.
\ee 
This integral can be evaluated numerically for any
choice of the parameters $b,c$ which give rise to a limit cycle. For
$b=2.2,c=1$ one finds $\mu_2=-0.20225$ along with the already
established observation that $\mu_1=0$.  The corresponding functions
$\bp^{(1)}(t)$ and $\bp^{(2)}(t)$ are illustrated in
Fig. \ref{fig:floquetCart}. In fact for the Brusselator the non-zero
exponent is bound to be real and negative throughout phases c) and d)
of the phase diagram (Fig. \ref{fig:pg}), i.e. throughout the limit
cycle phase.

In conclusion, we have established that one of the Floquet exponents of the 
system vanishes throughout this phase and that the remaining exponent assumes 
negative real values. The zero exponent is associated with perturbations in 
the longitudinal direction of the limit cycle; such perturbations are neither 
amplified nor reduced as the motion progresses. Perturbations in the transverse
direction, by contrast, decay in time in the Brusselator system, rendering the 
limit cycle stable. Indeed, the multiplier $\rho_2$ can be seen 
as characterizing a Poincar\'e map of transverse motion. If the system is 
perturbed transversely by a small amount, $\delta$, at time $t=0$ one may 
construct a Poincar\'e map in the usual way \cite{strogatz}: by forming the 
line perpendicular to the limit cycle which includes the point $\olx(t=0)$. 
Then at every integer multiple $nT$ of the period of the limit cycle the 
trajectory intersects the line at a distance $\rho_2^n\delta$ from 
the limit cycle. Since $\rho_2 < 1$, this approaches the limit cycle with 
increasing $n$.

%%%%%%%%%%%%%%%%%%%%%%%%%%%%%%%%%%%%%%%%%%%%%%%%%%%%%%%%%%%%%%%%%%%%%%%%%%%%%%

\subsection{Rotation into Frenet co-ordinates}\label{subsec:frenet}
\begin{figure}[t]
\centerline{\includegraphics[width=0.35\textwidth]{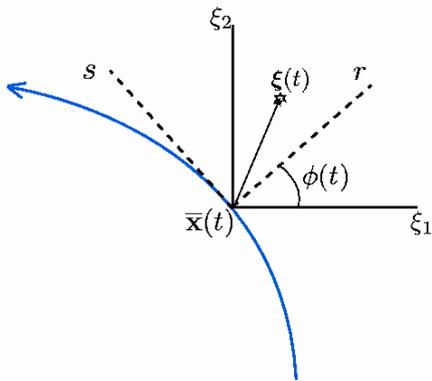}}
\caption{Illustration of the rotation from Cartesian co-ordinates 
$(\xi_1,\xi_2)$ to the Frenet frame, denoted by $(r,s)$. }
\label{fig:rotation}
\end{figure}
As seen above the Floquet exponents and periodic functions of the Brusselator 
system describe the relaxation of perturbations in longitudinal and transverse
directions. It is hence convenient to study the dynamics of deviations in 
co-ordinates defined along the tangential (longitudinal) and normal 
(transverse) directions to the limit cycle. This co-moving frame is generally 
referred to as the Frenet frame \cite{gibson}. We will label the transverse 
coordinate by $r$ and the longitudinal one by $s$, as illustrated in 
Fig. \ref{fig:rotation}. We follow the procedure of \cite{tomita} and define 
$\phi(t)$ to be the angle between the $x$-axis and the line normal to the 
limit cycle at time $t$. Transformation of perturbation displacement vectors 
from local Cartesian co-ordinates $\bxi=(\xi_1,\xi_2)$ into the co-moving 
Frenet frame $\bq\equiv(r,s)$ can then be thought of as a rotation by the angle
$\phi(t)$, i.e. $\bq(t)=J(t)\bxi(t)$, where
\BE
J(t) & = &\left(\begin{array}{cc}
\cos\phi(t) & \sin\phi(t)\\
-\sin\phi(t) & \cos\phi(t)\end{array}\right).
\EE
The angle $\phi(t)$, in turn, can be expressed as a
function of the velocity vector $\dot{\overline{\bx}}(t)$ at all points on the 
limit cycle trajectory. The transformation matrix is then
\BE
\label{eq:j}
J(t) & = &\frac{1}{v(t)}\left(\begin{array}{cc}
\dot{\overline{x}}_2(t) & -\dot{\overline{x}}_1(t)\\
\dot{\overline{x}}_1(t) & \dot{\overline{x}}_2(t)\end{array}\right),
\EE
where $v(t)$ is the speed (magnitude of the velocity) given by
$v(t)=\sqrt{(\dot{\overline{x}}_1)^2 + (\dot{\overline{x}}_2)^2}$. The time 
derivative of deviations from the limit cycle expressed in the Frenet frame 
is then given as
\begin{align}
\dot{\bq}(t) & =J(t)\dot{\bxi}(t)+\dot{J}(t)\bxi(t)\nonumber \\
& =\left(J(t)K(t)J^{-1}(t) + R(t)\right)\bq(t),
\label{eq:vectorDiffTrans}\end{align}
where $R(t)=\dot{J}(t)J^{-1}(t)$ has been introduced and explicitly we have,
\be
R(t) =\frac{\dot{\overline{x}}_2(t)\ddot{\overline{x}}_1(t)-
\dot{\overline{x}}_1(t)\ddot{\overline{x}}_2(t)}{v^{2}(t)}
\left(\begin{array}{cc}
0 & -1\\
1 & 0\end{array}\right).
\ee
To simplify notation we will use $K^\prime (t)\equiv J(t)K(t)J^{-1}(t)$ and 
$K^{\mathrm{tot}}(t)\equiv K^\prime (t) + R(t)$. As seen in \cite{tomita} the 
upper-right element of $K^\mathrm{tot}$ vanishes identically, i.e. we have 
$K^\prime_{rs}(t)+R_{rs}(t)=0$ for all times. Hence, the motion of the first 
co-ordinate $r(t)$ in the Frenet frame decouples from the second. More 
precisely, Eq. (\ref{eq:floquet}) takes the form
\BE\label{eq:rs}
\frac{\mathrm{d}}{\mathrm{d}t}\left(\begin{array}{c}
r(t)\\
s(t)\end{array}\right)=\left(\begin{array}{cc}
K_{rr}^{\mathrm{tot}}(t) & 0\\
K_{sr}^{\mathrm{tot}}(t) & K_{ss}^\mathrm{tot}(t)\end{array}\right)
\left(\begin{array}{c}
r(t)\\
s(t)\end{array}\right)
\EE
after rotation into co-moving co-ordinates. The non-trivial elements of 
$K^{\mathrm{tot}}$ can be computed explicitly as functions of the limit cycle 
trajectory and are given by
\BE
K_{rr}^{\mathrm{tot}}&=&\frac{1}{v^2}\big\{\olxonedot^2K_{22}+
\olxtwodot^2K_{11}\nonumber\\
&&~~~~~~-\olxonedot\olxtwodot(K_{12}+K_{21})\big\}, \\
K_{sr}^{\mathrm{tot}}&=&\frac{1}{v^2}\big\{(\olxtwodot^2-\olxonedot^2)
(K_{12}+K_{21})\nonumber\\
&&~~~~~~+2\olxonedot\olxtwodot(K_{11}-K_{22})\big\}, \\
K_{ss}^{\mathrm{tot}}&=&\dot v/v.\label{eq:vdotbyv}
\EE
%%%%%%%%%%%%%%%%%%%%%%%%%%%%%%%%%%%%%%%%%%%%%%%%%%%%%%%%%%%%%%%%%%%%%%%%%%%%%%
\subsection{Re-scaled Frenet frame}\label{subsec:frenet_scaled}
\begin{figure}
\includegraphics[clip,width=1\columnwidth,keepaspectratio]
{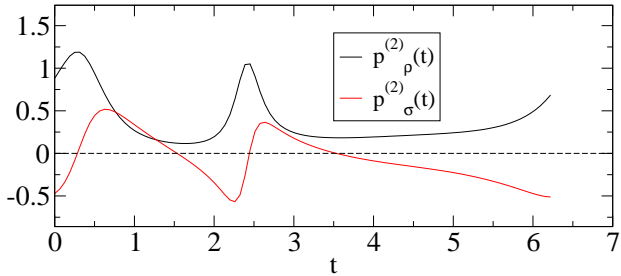}
\caption{\label{fig:floquetTau} (Color on-line) Periodic functions 
$p_\rho^{(2)}(t)$ and $p_\sigma^{(2)}(t)$  associated with the non-trivial 
Floquet multiplier $\rho_2$ of Eq. (\ref{eq:linearTau}). While a perturbation 
$\rho$ decays in a non-linear fashion, the phase, $\sigma$, oscillates.}
\end{figure}

Further simplification can be achieved by re-scaling the co-ordinates of the 
Frenet frame, after rotation, by the velocity of the limit cycle. More 
precisely we make the transformation
$\mathbf{\bchi}(t)\equiv \bq (t)/v(t)$, so that 
\be
\label{eq:lambdadef}
\mathbf{\bchi}(t) = \frac{1}{v(t)}J(t)\bxi(t)\equiv \Lambda(t)\bxi(t).
\ee
We will denote the individual components by $\bchi(t)=(\rho(t),\sigma(t))$. 
One finds that the perturbative displacement is described by
\be\label{eq:chidot}
\dot \bchi(t)=\left( K^{\mathrm{tot}}-\frac{\dot v(t)}{v} I \right) \bchi (t)
\equiv L^{\mathrm{tot}}\bchi (t),
\ee
where $I$ denotes the identity matrix. Combining Eqs. (\ref{eq:rs}) and 
(\ref{eq:vdotbyv}), Eq. (\ref{eq:chidot}) takes the simple form,
\begin{equation}
\frac{\mathrm{d}}{\mathrm{d}t}\left(\begin{array}{c} \rho(t)\\
\sigma(t)\end{array}\right)=\left(\begin{array}{cc}
L_{\rho\rho}^{\mathrm{tot}}(t) & 0\\ L_{\sigma\rho}^{\mathrm{tot}}(t)
& 0\end{array}\right)\left(\begin{array}{c} \rho(t)\\
\sigma(t)\end{array}\right).\label{eq:linearTau}
\end{equation} In
particular the lower right element of $L^{\mathrm{tot}}$ vanishes
identically. The remaining elements are given by 
\BE
L_{\rho\rho}^{\mathrm{tot}}(t) &=&\frac{1}{v^2}
\big\{(\olxtwodot^2-\olxonedot^2)(K_{11}-K_{22})\nonumber\\
&&~~~~~~-2\olxonedot\olxtwodot(K_{12}+K_{21})\big\}, \\
L_{\sigma\rho}^{\mathrm{tot}}(t) &=&\frac{1}{v^2}
\big\{(\olxtwodot^2-\olxonedot^2)(K_{12}+K_{21})\nonumber\\
&&~~~~~~+2\olxonedot\olxtwodot(K_{11}-K_{22})\big\}.  
\EE 
These relations are valid for general systems of first-order
ordinary differential equations with two degrees of freedom which exhibit
limit cycle solutions. The vanishing elements in Eq. (\ref{eq:linearTau}) 
guarantee the existence of the constant solution $\bchi(t)\equiv(0,\sigma_0)$ 
(with $\sigma_0$ a real-valued constant). Hence we find that a perturbation in 
$\sigma$ is not only periodic, as indicated by the trivial Floquet exponent,
but it is in fact constant. Transverse perturbations relax in a non-trivial 
way and for completeness we show the associated periodic functions, 
$p_\rho^{(2)}(t)$ and $p_\sigma^{(2)}(t)$ in Figure \ref{fig:floquetTau}. The 
superscript here indicates that we refer to the non-trivial Floquet multiplier 
$\rho_2$. It should be noted that there is a secondary oscillatory effect on 
$\sigma$ as $\rho$ decays back to the limit cycle. Since this 
velocity-scaled rotation yields a simpler linear theory than rotation alone, 
we will only use the co-ordinates $(\rho,\sigma)$ in the discussions that 
follow.

%%%%%%%%%%%%%%%%%%%%%%%%%%%%%%%%%%%%%%%%%%%%%%%%%%%%%%%%%%%%%%%%%%%%%%%%%%%
%%%%%%%%%%%%%%%%%%%%%%%%%%%%%%%%%%%%%%%%%%%%%%%%%%%%%%%%%%%%%%%%%%%%%%%%%%%
\section{Stochastic effects in the limit cycle phase}\label{sec:kampen}
\subsection{System size expansion and analytical predictions}\label{subsec:SSE}
We will now return to the stochastic system defined by the chemical
reactions (\ref{eq:reactions}) or, equivalently, by the master equation
(\ref{eq:Masterequation}). In this section we will apply van Kampen's 
system-size expansion to the case where we have a limit cycle, 
$(\overline{x}_1(t),\overline{x}_2(t))$, in the mean-field. The starting point 
is the transformation of random variables $\bn\mapsto \bxi(t)$ defined by
\be 
\bn=N\olx(t)+\sqrt{N}\bxi(t).\label{eq:vkChangeofvariable} 
\ee
The only difference compared to the transformation
(\ref{eq:SSE_FixedPoint}) we applied in the fixed-point phase is the
time dependence of the first term on the right-hand side. Apart from
this minor complication the algebraic steps necessary to carry out the
expansion of the master equation are mostly unchanged, see the
Appendix for details. As before, the evolution of the stochastic
fluctuations is described by a linear Langevin equation, 
\BE
\dot{\bm{\xi}}(t)=K (t)\bm{\xi}+\mathbf{f}(t)\label{eq:Langevincycle},
\EE 
where $\mathbf{f}(t)=(f_1(t),f_2(t))$ is bivariate Gaussian white
noise with zero mean with correlations given by,
$\avg{f_i(t)f_j(t')}=2D_{ij}(t)\delta(t-t')$ \cite{tomita,scott}. The
forms of the matrices $K(t)$ and $D(t)$ are given in the Appendix
(Eqs. (\ref{eq:Kform}) and (\ref{eq:Dform}), respectively). Because of the
periodicity of $\overline{\mathbf{x}}(t)$, all elements of $K(t)$ and
$D(t)$ are periodic with the period of the limit cycle, $T$.

The linearity of these Langevin equations allows us to make further analytical 
progress. To this end, it is convenient to study the stochastic dynamics 
in the velocity-scaled Frenet frame as introduced above. Upon performing a 
rotation into the co-ordinate system spanned by $(\rho,\sigma)$, 
Eq. (\ref{eq:Langevincycle}) takes the form
\BE
\frac{\mathrm{d}}{\mathrm{d}t}\left(\begin{array}{c}
\rho(t)\\
\sigma(t)\end{array}\right)=\left(\begin{array}{cc}
L_{\rho\rho}^{\mathrm{tot}}(t) & 0\\
L_{\sigma\rho}^{\mathrm{tot}}(t) & 0\end{array}\right)\left(\begin{array}{c}
\rho(t)\\
\sigma(t)\end{array}\right)+\left(\begin{array}{c}
\zeta_1(t)\\
\zeta_2(t)\end{array}\right), \nonumber \\
\label{eq:Langevincyclerhosigma}
\EE
with $\langle \zeta_{i} (t) \rangle = 0,\ i=1,2$ and 
\be
\label{eq:zetacorrel}
\avg{\zeta_i(t)\zeta_j(t')}=2[\Lambda(t) D(t)\Lambda^\mathrm{T}(t)]_{ij}
\delta(t-t').
\ee
Hence the correlations of the noise components in the co-moving frame are 
described by the matrix $H(t)\equiv \Lambda(t) D(t)\Lambda^\mathrm{T}(t)$,
where $\Lambda(t)$ is defined by Eq. (\ref{eq:lambdadef}).

\begin{figure}[t]
\centerline{\includegraphics[width=.7\columnwidth]{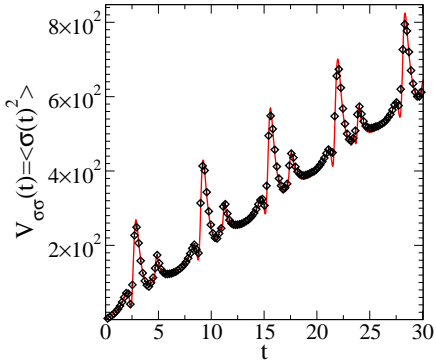}}
\caption{(Color on-line) Variance $V_{\sigma\sigma}(t)$ of longitudinal 
fluctuations as obtained from 
solving Eqs. (\ref{eq:variance1})-(\ref{eq:variance3}). Model parameters are 
fixed to $b=2.2,c=1$. The solid line is calculated from the theory while the 
markers are obtained from simulations. System size is $N=10^5$, averages over 
$10000$ runs are taken.}
\label{fig:varth}
\end{figure}

Since the mean values of both $\rho$ and $\sigma$ vanish, one can 
write the variances of these variables as 
$V_{\rho\rho}(t)=\left\langle \rho^2(t)\right\rangle$ and 
$V_{\sigma\sigma}(t)=\left\langle \sigma^2(t)\right\rangle$. First-order 
ordinary differential equations can be derived \cite{vankampen} for these 
quantities, and they take the form
\BE
\dot{V}_{\rho\rho}(t)&=&2L^\mathrm{tot}_{\rho\rho}V_{\rho\rho}(t)
+2H_{\rho\rho}(t)\label{eq:variance1},\\
\dot{V}_{\sigma \rho}(t)&=&L^\mathrm{tot}_{\sigma \rho}V_{\rho\rho}(t)
+L^\mathrm{tot}_{\rho\rho}V_{\sigma \rho}(t)+
2H_{\sigma \rho}(t),\\
\dot{V}_{\sigma\sigma}(t)&=&2L^\mathrm{tot}_{\sigma \rho}V_{\sigma\rho}(t)
+2H_{\sigma\sigma}(t).\label{eq:variance3}
\EE
Solving Eqs (\ref{eq:variance1})-(\ref{eq:variance3}) sequentially, a closed 
form for $V_{\sigma\sigma}(t)$ may be found and evaluated numerically. Results 
are shown in Fig. \ref{fig:varth}. Note that the variance of $\sigma$, 
evaluated at integer multiples of the time period, $T$, increases linearly. 
This increase without bound means that the linear approximation within which 
we derived our theoretical results can be expected to be valid only at 
sufficiently short times. More precisely the first-order van Kampen expansion 
is accurate provided $\sigma(t)/\sqrt{N}$ is small compared to the components 
of the limit cycle solution, $\olx(t)$. The time scale on which longitudinal 
fluctuations remain small enough for the linear theory to be valid will hence 
increase as the system size is increased, and will diverge as $N\to\infty$.

The Langevin equation for radial fluctuations completely decouples from that 
of $\sigma$. From Eq. ({\ref{eq:Langevincyclerhosigma}}) we have,
\be\label{eq:langevinsimple}
\dot \rho(t)=L_{\rho\rho}^\mathrm{tot}(t)\rho(t)+\zeta_1(t),
\ee
which is readily integrated, given an initial condition $\rho(t_0)=\rho_0$:
\be
\rho(t)-\rho_0\Phi(t,t_0)=\int_{t_0}^t \Phi(t,t')\zeta_1(t')\mathrm{d}t',
\label{eq:integraterho}
\ee
where we have used the definition
\be
\Phi(t,t')=\exp\left(\int_{t'}^t  L_{\rho\rho}^\mathrm{tot}(t'')
\mathrm{d}t''\right).
\ee
We can now evaluate the average temporal correlations of radial fluctuations 
for $\tau > 0$:
\BE
&&\avg{\rho(t+\tau)\rho(t)}-\rho_0\Phi(t+\tau,t)\Phi^2(t,t_0)\nonumber\\
&=&2\int_{t_0}^t\Phi(t+\tau,t')\Phi(t,t') H_{\rho\rho}(t')  \mathrm{d}t'
\nonumber\\
&=&2\Phi(t+\tau,t)\int_{t_0}^{t}\Phi^2(t,t^{\prime})
H_{\rho\rho}(t^{\prime})\mathrm{d}t',
\EE
where we have used the identity $\phi(t,t')\phi(t',t'')=\phi(t,t'')$, valid 
for all  $t,t',t''$. We now set the initialization time to the infinite past 
$(t_0\to -\infty)$ so that the initial condition itself is forgotten. After 
invoking the periodicity of $L_{\rho\rho}(t)$ and $H_{\rho\rho}(t)$ and making 
a suitable change of the integration variable one finds
\BE
\avg{\rho(t+\tau)\rho(t)}&=&\frac{2\Phi(t+\tau,t)}{1-\mathrm{e}^{2\mu_\rho}}
\nonumber \\
&&\hspace{-5em}\times \int_{0}^{T}\Phi^2(t+T,t+t^{\prime})H_{\rho\rho}
(t+t^{\prime})\mathrm{d}t',\label{eq:rttau}
\EE
where the pre-factor is due to an infinite summation of powers of 
$\mathrm{e}^{2\mu_\rho}$, resulting from the observation that 
$\Phi(t+T,t)=e^{\mu_\rho}$, where $\mu_\rho$ is the non-vanishing Floquet 
exponent of the system. Eq. (\ref{eq:rttau}) confirms that the two-time 
correlation $\avg{\rho(t+\tau)\rho(t)}$ is periodic in $t$. Averaging over 
$t$ yields the (time-averaged) autocorrelation function 
\cite{gardiner:handbook,riley},
\BE
C(\tau)\equiv\frac{1}{T}\int_{0}^{T} 
\left\langle \rho(t)\rho(t+\tau)\right\rangle\mathrm{d}t.
\EE
The power spectrum of $\rho(t)$ is now obtained as the inverse Fourier 
transform of this function $C(\tau)$. A similar procedure can be implemented 
to compute the power spectrum of $r(t)$, the only necessary replacements being 
$H_{\rho\rho}(t)\to G_{rr}(t) \equiv [J(t)D(t)J^\mathrm{T}(t)]_{rr}$ and 
$L_{\rho\rho}^\mathrm{tot}(t)\to K_{rr}^\mathrm{tot}(t)$.

\begin{figure}[t]
\centerline{\includegraphics[width=0.35\textwidth]{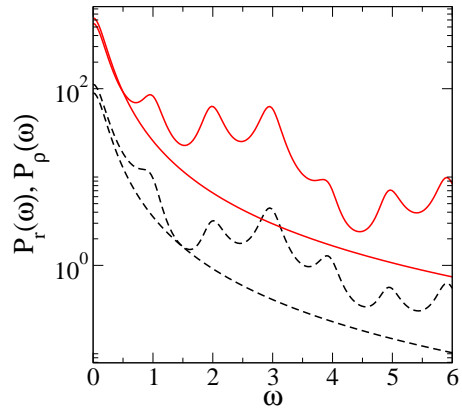}}
\caption{(Color on-line) Power spectra $P_\rho(\omega)$ and $P_r(\omega)$ of 
transverse fluctuations in the velocity scaled Frenet frame and ordinary Frenet
frame for the Brusselator at $c=1$ and $b=2.2$. The solid pair of curves shows 
results for $P_\rho(\omega)$ while the dashed-line pair shows those without 
the rescaling of the Frenet frame. The monotonic curves in each case correspond
to Lorentzians obtained by replacing 
$L_{\rho\rho}^\mathrm{tot}(t), H_{\rho\rho}(t)$ and 
$K_{rr}^\mathrm{tot}, G_{rr}(t)$ by their time averages in 
Eq. (\ref{eq:rttau}).}
\label{fig:spectanalytical}
\end{figure}

The resulting spectra of transverse fluctuations are plotted in
Fig. \ref{fig:spectanalytical}, both in the rescaled Frenet frame (the
power spectrum is then denoted by $P_\rho(\omega)$), and in the
ordinary Frenet frame, see the curve referred to as $P_r(\omega)$. Peaks of a 
finite width are found at multiples of the frequency of the limit cycle and it 
can be seen that the structure of these peaks is dependent only on the 
parameters $b$ and $c$ and not on system size, $N$. For comparison we also 
show the power spectra one would obtain by replacing the periodic matrices 
$L^\mathrm{tot}(t)$, $K^\mathrm{tot}(t)$, $H(t)$ and $G(t)$ by their mean 
values (e.g. $\overline{H}\equiv T^{-1}\int_0^T H(t)\mathrm{d}t$) in 
Eq. (\ref{eq:rttau}). In this case the power spectra reduce to Lorentzian 
curves which we plot in Fig. \ref{fig:spectanalytical} along with the full 
results. It would seem a plausible conclusion that the time-dependence of the 
drift and diffusion matrices ($L^\mathrm{tot}(t)$ and $H(t)$, respectively) 
contributes additively to the Lorentzian shape that we would expect from a 
Langevin equation with time-independent drift and diffusion matrices.

%%%%%%%%%%%%%%%%%%%%%%%%%%%%%%%%%%%%%%%%%%%%%%%%%%%%%%%%%%%%%

\subsection{Test against numerical simulations}\label{subsec:numerical}
\begin{figure}[t]
\centerline{\includegraphics[width=0.35\textwidth]{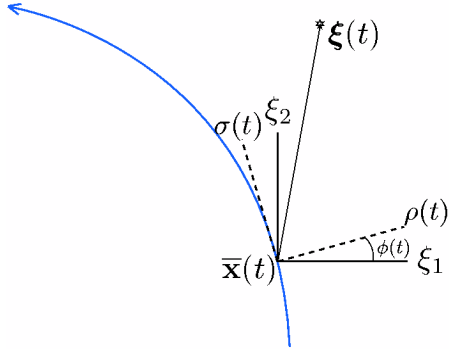}}
\caption{Illustration of how data is obtained from Gillespie simulations. At 
any time $t$ the algorithm generates a point $(n_1(t)/N,n_2(t)/N)$ in the 
$x_1,x_2$-plane. The vector components of the vector $\bxi(t)$ are then 
obtained as $\xi_1(t)= N^{1/2}\left[n_1(t)/N-\olxone(t)\right]$ and similarly 
for $\xi_2(t)$. As before, $\overline{\bx}(t)$ is the point marked by the 
limit cycle trajectory at time $t$. Subsequently, $\bxi(t)$ is converted into 
co-ordinates $\bchi(t)=(\rho(t),\sigma(t))$ by the transformation 
$\bchi(t)=\Lambda(t)\bxi(t)$, with $\Lambda(t)$ as defined in 
Eq. (\ref{eq:lambdadef}).}
\label{fig:gillInt1}
\end{figure}
The above theoretical prediction for the power spectrum of transverse 
fluctuations about the mean field behavior can be verified in simulations of 
{\em finite} systems using the well known Gillespie algorithm \cite{gill} to 
simulate the reaction system of the Brusselator model. This method generates 
continuous-time realizations $(n_1(t),n_2(t))$ of the multi-particle stochastic
process described by the master equation (\ref{eq:Masterequation}). Note that 
$n_1(t)$ and $n_2(t)$ are integer valued at all times. The overall system size,
$N$, is a control parameter in these simulations, as are the reactivity 
parameters $b$ and $c$.

To make contact with the above theoretical analysis we begin by considering 
the transformation $\bn\mapsto\bxi$ given in Eq. (\ref{eq:vkChangeofvariable}),
i.e. we have,
\BE
n_1(t)/N&=&\olxone(t)+\xi_1(t)/N^{1/2}, \nonumber \\
n_2(t)/N&=&\olxtwo(t)+\xi_2(t)/N^{1/2}. \label{eq:gillexpansion} 
\EE
The quantities $\rho$ and $\sigma$ are then obtained by performing the 
transformation 
$(\rho(t),\sigma(t))^\mathrm{T}=\Lambda(t)(\xi_1(t),\xi_2(t))^\mathrm{T}$, 
with $\Lambda(t)$ as defined in Eq. (\ref{eq:lambdadef}). This is illustrated 
in Fig. \ref{fig:gillInt1}.

\begin{figure}[t]
\centerline{\includegraphics[width=0.35\textwidth]{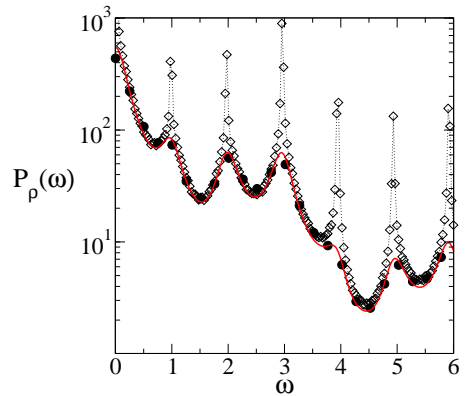}}
\caption{(Color on-line) Power spectrum $P_{\rho}(\omega)$ of transverse 
fluctuations about the limit cycle trajectory. Model parameters are 
$c=1,b=2.2$. The solid curve results from theoretical analysis while filled 
symbols are from Gillespie simulations run up to $t_\mathrm{f}=25$ and 
connected open markers are from simulations run up to $t_\mathrm{f}=200$. 
Simulations interpreted using the rotation method as explained in 
Section \ref{subsec:numerical}. Averages over $10000$ samples are 
taken, system size is $N=10^5$.}
\label{fig:spect1}
\end{figure}

The linear theory we have developed can be expected to be accurate at most in 
a regime where the second terms on the right-hand sides of 
Eq. (\ref{eq:gillexpansion}) are both small corrections to the first terms. 
Equivalently we require that $\rho(t)$ and $\sigma(t)$ both remain small 
compared to $N^{1/2}$. However, as we have already discussed and demonstrated 
in Fig. \ref{fig:varth}, the variance of $\sigma(t)$ grows linearly in time 
(modulo periodic variations). Hence the results based on the van Kampen 
expansion are expected to be accurate at most on time scales 
$t_\mathrm{f}={\cal O}(N^0)$. Performing simulations at system size $N$ and 
run up to time scales $t_\mathrm{f}\ll N$ one observes good agreement with the 
predictions of the linear theory  as illustrated by Fig. \ref{fig:spect1}, 
where we show data for systems of size $N=10^5$ run up to $t_\mathrm{f}=25$. 
However, when we consider simulations run up to larger times 
(e.g. $t_\mathrm{f}=200$), the power spectrum shows systematic deviations from 
the theoretical curve. This effect can be accounted for by the reasonably 
steep average slope found in Fig. \ref{fig:varth}. Extrapolating 
Fig. \ref{fig:varth} to larger times one expects 
$\avg{\sigma(t=200)^2}\approx 4000$, 
i.e. $\left[\avg{(\sigma(t=200)/\sqrt{N})^2}\right]^{1/2}\approx 0.2$ for 
$N=10^5$, so that the second terms on the right-hand sides of 
Eq. (\ref{eq:gillexpansion}) can no longer be thought of as small compared to 
the first.  The next section will discuss an alternative set of measurements 
that may be taken from Gillespie simulations. These will be shown to 
successfully tackle this defect and yield a good match with the prediction 
illustrated in Fig. \ref{fig:spectanalytical} also on longer time scales.
%%%%%%%%%%%%%%%%%%%%%%%%%%%%%%%%%%%%%%%%%%%%%%%%%%%%%%%%%%%%%%%%%%

\subsection{Modified method of comparison}\label{subsec:modified}
\begin{figure}[t]
\centerline{\includegraphics[width=0.25\textwidth]{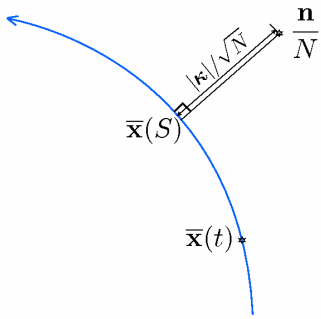}}
\caption{The modified procedure of identifying deviations from the limit cycle 
trajectory.  The point of reference is now the point on the limit cycle 
trajectory, which is geometrically closest to the point in the Cartesian plane 
obtained in Gillespie simulations. }
\label{fig:gillInt2}
\end{figure}

The approach that we took in the previous section to interpreting time series 
from the Gillespie simulations compares extremely well with the theory we have 
developed in terms of the variance of $\sigma$. However, except for simulations
performed on very short time scales, we find that this is not the case for the 
power spectrum of $\rho$. We will therefore pursue an alternative way of 
comparing data from Gillespie simulations with the power spectra obtained from 
the theory. Specifically, the proposal put forward in this section is to use a 
transformation $\bn\mapsto\bkappa$ of the form
\be
\bn=N\olx(S)+\sqrt{N}\bkappa (\bn, S),
\label{mod_1}
\ee
where $S$ is a random variable chosen as 
\be
S\equiv\mbox{arg min}_{t'} |N\olx(t')-\bn|.
\label{mod_2}
\ee
Here $\mbox{arg min} |N\olx(t')-\bn|$ denotes the value $t'$ minimizing the 
quantity $|N\olx(t')-\bn|$, i.e. $S$ is chosen such that $\olx(S)$ is the point
on the limit cycle trajectory with the minimum distance to the point $\bn/N$ 
obtained from the Gillespie simulation ($|\cdot|$ refers to the Euclidean 
norm). By construction the vector $\bkappa$ is then perpendicular to the 
velocity of the limit cycle: $\bkappa(\bn, S).\dot{\olx}(S)=0$. In this way we 
are able to define a Frenet frame directly in terms of the variable $\bn$ 
which appears in the original master equation, rather than constructing it as 
a rotation from continuous Cartesian coordinates.

Although our intention is to use this formulation to reinterpret the simulation
data, we first have to show how the above construction recovers the results
obtained through the van Kampen system-size expansion. Since $S$ is a function
of the stochastic variable $\bn$, we may define its mean value as
\be
\langle S \rangle_t = \sum_{\bn} S(\bn) P_{\bn} (t),
\label{S_av}
\ee
where $P_{\bn}(t)$ is the solution of the master equation. Fluctuations 
about this mean value are defined by
\be
S(\bn) = \langle S \rangle_t + \frac{1}{\sqrt{N}} \sigma(\bn, t).
\label{S_fluct}
\ee 
Neglecting terms of order $1/N$, as before, we may Taylor expand about the
point $\langle S \rangle_t$ on the curve to obtain 
\be
\olx(S) = \olx(\langle S \rangle_t) + \frac{1}{\sqrt{N}} \sigma(\bn, t) 
\dot{\olx}(t).
\label{Taylor}
\ee
Using Eqs. (\ref{mod_1}) and (\ref{Taylor}) we have that
\be
\bn=N\olx(\langle S \rangle_t)+\sqrt{N} \left\{ \sigma(\bn, t) \dot{\olx} (t)
+ \bkappa (\bn, S) \right\}.
\label{mod_VK}
\ee
To the order we are working, we may replace $\bn$ and $S$ in the curly 
bracket by $N\olx (\langle S \rangle_t)$ and $\langle S \rangle_t$ 
respectively, by using Eqs. (\ref{S_fluct}) and (\ref{mod_VK}). Furthermore 
from Eq. (\ref{mod_VK}) we have $\bn = N \olx(\langle S \rangle_t)$ to leading 
order. Since the limit cycle is defined by $\olx$, the correct mean-field 
equations are only recovered if the identification $\langle S \rangle_t = t$ is
made. These considerations lead to Eq. (\ref{mod_VK}) being written as
\be
\bn=N\olx(t)+\sqrt{N} \left\{ \sigma(t) \dot{\olx} (t) + \bkappa (t) \right\}.
\label{mod_VK_2}
\ee
Comparing Eq. (\ref{mod_VK_2}) with the usual starting point for the van 
Kampen expansion, Eq. (\ref{eq:vkChangeofvariable}), we see that for these two
approaches to agree it must be the case that
\be
\bxi(t) = \sigma(t) \dot{\olx} (t) + \bkappa (t).
\label{ident}
\ee
This is so, since to leading order the condition 
$\bkappa(\bn, S).\dot{\olx}(S)=0$ becomes $\bkappa(t).\dot{\olx}(t)=0$. 
Therefore from Eq. (\ref{ident}) we require that 
$\sigma (t)= \bxi(t).\dot{\olx}(t)/v^{2}(t)$. But from $\bq(t)=J(t)\bxi(t)$ and
Eq. (\ref{eq:j}) it follows that $s(t) =\bxi(t).\dot{\olx}(t)/v(t)$ and
from Eq. (\ref{eq:lambdadef}) that $\sigma(t) = s(t)/v(t)$. This
identifies $\sigma$ as the scaled, longitudinal component introduced in 
Section \ref{subsec:frenet_scaled}. By construction 
$\bkappa = \bxi - \sigma \dot{\olx}$ must be the transverse component. In this 
way, we recover the van Kampen ansatz used previously.

\begin{figure}[t]
\centerline{\includegraphics[width=0.35\textwidth]{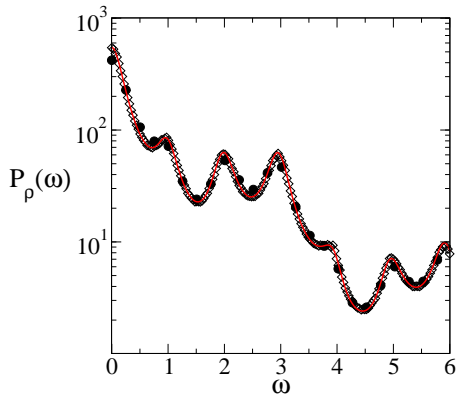}}
\caption{(Color on-line) Power spectrum $P_{\rho}(\omega)$ of transverse 
fluctuations about the limit cycle trajectory. Model parameters are 
$c=1,b=2.2$. The solid curve results from theoretical analysis while filled 
symbols are from Gillespie simulations run up to $t_\mathrm{f}=25$ and 
connected open markers are from simulations run up to $t_\mathrm{f}=200$. 
Simulations are interpreted using the projection method as explained in 
Section \ref{subsec:modified}. Averages over $10000$ samples are taken, system 
size is $N=10^5$.}
\label{fig:spect2}
\end{figure}

The Gillespie algorithm generates a time-series for $\bn$ which we can use 
together with knowledge of the equation for the limit cycle to determine $S$,
from Eq. (\ref{mod_2}). This can then be used (i) to determine $\sigma$, which
by Eq. (\ref{S_fluct}) and the identification $\langle S \rangle_t = t$ is 
$\sqrt{N} (S-t)$, (ii) to determine $\bkappa$ from Eq. (\ref{mod_1}), and 
(iii) to obtain re-scaled transverse fluctuations 
$\rho(t)=\psi(t)|\bkappa(t)|/v(S)$. The function $\psi(t)$ is present to ensure
the correct sign; we choose $\psi(t)=-1$ whenever $\bn$ is located inside the 
limit cycle trajectory, and $\psi(t)=1$ if $\bn$ lies outside the area 
surrounded by the limit cycle. It should be noted that now time is introduced 
only from the Gillespie algorithm, and not through Eq. (\ref{mod_1}). Using 
this methodology, we may determine the power spectra from repeated Gillespie 
simulations. The results for the variance of $\sigma$ are indistinguishable 
from those shown in Figure \ref{fig:varth} and so we do not show them. However,
in Fig. \ref{fig:spect2} we compare the power spectrum estimated by measuring 
$\rho(t)=\psi(t)|\bkappa(t)|/v(S)$ from simulations with the analytical 
results obtained in Section {\ref{subsec:SSE}} (see 
Eq. (\ref{eq:langevinsimple})). As seen in the figure, we find good 
quantitative agreement also on time scales on which the method discussed in 
the previous section failed to reproduce the theoretical curve. 

The findings of the section may be summarized by postulating three different
temporal regimes. The first regime is defined for times which are sufficiently
short that $\sigma/\sqrt{N}$ can be thought of as small compared with the
size of the limit cycle. Thus Eq. (\ref{eq:gillexpansion}) may be used, and
the modified approach based on Eq. (\ref{mod_1}) need not be used. The results
for $t_\mathrm{f}=25$ exemplify this regime. In the second regime, 
$\sigma/\sqrt{N}$ is now sufficiently large that the use of the naive 
expression (\ref{eq:gillexpansion}) leads to a disagreement between simulations
and the theoretical curve. This has been discussed above and is exemplified 
by the results for $t_\mathrm{f}=200$. Finally, at longer times $\sigma$ will
start to probe the periodic structure of the limit cycle, and a different
type of behavior will occur. We have not explored this latter regime in the 
present paper, but we will discuss it again in Section \ref{sec:discussion}.

\begin{figure}[t]
\vspace{2em}
\centerline{\includegraphics[width=0.5\textwidth]{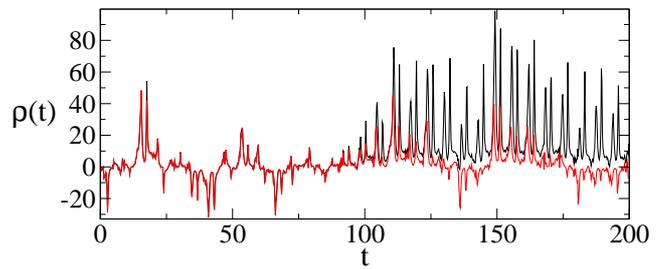}}
\vspace{3em}
\caption{(Color on-line) Single time series of transverse fluctuations, 
$\rho(t)$, illustrating the difference between the two measurement methods. 
The black curve shows data obtained using the rotation method (see Section
\ref{subsec:numerical}), the red curve shows result from the projection method 
(see Section \ref{subsec:modified}). Data is obtained from a single Gillespie 
run at $N=10^5$ ($b=2.2$, $c=1$).}
\label{fig:timeseries}
\end{figure}
We end with a further illustration of the difference between the two methods 
of extracting transverse fluctuations from simulation data. We show an example 
of a single run in Fig. \ref{fig:timeseries} and depict the resulting two time 
series $\rho(t)$ produced using the two different methods. Initially the two 
time series agree well with one another. On longer time scales, however, 
systematic deviations are observed. We here note that the shape of time series 
as shown in Fig. \ref{fig:timeseries} can vary considerably between different 
runs of the stochastic Gillespie simulations. For reasons of clarity the 
realization shown in the figure is one where the deviation between the time 
series generated using the two methods is reasonably pronounced at large
times. In other runs the discrepancy was smaller.

%%%%%%%%%%%%%%%%%%%%%%%%%%%%%%%%%%%%%%%%%%%%%%%%%%%%%%%%%%%%%%%%%%%%%%%%%%%%%%
%%%%%%%%%%%%%%%%%%%%%%%%%%%%%%%%%%%%%%%%%%%%%%%%%%%%%%%%%%%%%%%%%%%%%%%%%%%%%%
\section{Discussion}\label{sec:discussion}
In summary we have carried out an analysis of the effects of internal 
fluctuations found in finite systems with a large system size, $N$. We have 
focused on two-dimensional systems and we have used the Brusselator 
model as a toy example which we have discussed in two regimes. In the case 
where the mean-field dynamics approaches a stable fixed-point behavior with 
complex eigenvalues, we found the expected sustained oscillations driven by 
the stochasticity of the discrete particle dynamics. The power spectra of 
these oscillations can be obtained analytically via an expansion in the 
inverse system-size about the time-independent solution. This is similar to 
work carried out in the context of other models with fixed-point 
behavior \cite{alan1,alan2,lama,kuske}. 

One of the aims of the present work was to extend these analytical tools 
to the case in which the dynamics exhibit a periodic solution on the mean-field
level for a range of parameter values. The above van Kampen expansion in the 
inverse system size can be carried out as before. However, now that we are 
expanding about a curve, rather than a point, we must make a choice for the
point on the limit cycle about which we expand. This scheme may be left 
undetermined for carrying out the system-size expansion. However when carrying 
out simulations, more care has to be exercised when picking the point that one
expands about. At short times, the time since the start of the simulation can
be used to determine the point along the limit cycle. However, at longer times 
this will typically not coincide with the nearest point to the quantity 
$\bn/N$, due to the diffusion in the direction tangential to the limit cycle.
We have given a prescription for carrying out simulations in the second 
temporal regime, and shown how we recover agreement between simulations and 
the van Kampen expansion using this method. As we pointed out, there is a 
third regime where $\sigma/\sqrt{N}$ becomes of the order of the period $T$,
when further modifications will have to be introduced. Eventually, on very
long timescales, the diffusive behavior can be studied by using $N^{-1}$ 
rather than $N^{-1/2}$ as an expansion parameter \cite{vankampen}. It would
be interesting to extend the work we have presented here to these longer 
times.

Another objective was to understand the relation between the cycles due to the
stochastic amplification in the fixed point phase and the limit cycles in the 
phase where the fixed point becomes unstable. We have shown how, for the
example of the Brusselator, the former become the latter as one passes through 
the phase boundary. In the fixed point phase the fluctuations are amplified by
a resonance which may be described by a pole in the complex frequency plane.
As the phase boundary is approached the pole migrates towards the real axis,
reaching it when the boundary is crossed and so turning the resonance into a
limit cycle. Although this has been illustrated in the case of the Brusselator,
we expect this phenomenon to be generic, and that it may be applied to the 
various systems mentioned in the Introduction.

The output of the Gillespie algorithm is a time series similar to that which 
is found in data obtained from real systems. We therefore expect that the
methods we have applied in this paper will be applicable to real data. We hope
that this will lead to further insights when applied to the many other systems 
which have a variety of stable attractors and which are subject to intrinsic 
noise. 

\begin{acknowledgments}
RPB would like to thank EPSRC for the award of a postgraduate
grant. TG is an RCUK Fellow (RCUK reference EP/E500048/1). 
\end{acknowledgments}

\renewcommand{\theequation}{A-\arabic{equation}}
% redefine the command that creates the equation no.
\setcounter{equation}{0}  % reset counter 
\appendix
\section*{Appendix: Van Kampen System Size Expansion}
In this appendix we briefly sketch some of the mathematical steps involved in
carrying out the system-size expansion for two-dimensional chemical systems,
of which the Brusselator is an example. The starting point is the master 
equation (\ref{eq:Masterequation}),
\BE\label{eq:masterapp}
\frac{\mathrm{d} P_{\mathbf{n}}(t)}{\mathrm{d}t}&=&
\sum_{\nu}\left(T_{\nu}(\mathbf{n}-\mathbf{v}_{\nu})
P_{\mathbf{n}-\mathbf{v_{\nu}}}(t)
-T_{\nu}(\mathbf{n})P_{\mathbf{n}}(t)\right),\nonumber
\\
\EE
which, subject to an initial condition $P_\bn(t_0)$, governs the temporal 
evolution of the probability distribution describing the statistics of the 
microscopic dynamics as defined by the reactions (\ref{eq:reactions}). 
%Anticipating that the distribution
% of state variables $n_1(t)/N$ and $n_2(t)/N$ at any time $t$ will be
% of a Gaussian shape with mean given by the mean-field trajectory, i.e.
% $\avg{\mathbf{n}(t)}/N= (x_1(t),x_2(t))$, we will write
We determine in Section \ref{sec:kampen} that we may use the following
mapping between the population vector $\bn$ and a continuous fluctuation, 
$\bxi(t)$, about the mean-field concentration as follows,
\be
\mathbf{n}=N\bx(t)+\sqrt{N}\bxi(t).
\ee
Following the procedure of \cite{vankampen} we now formulate the problem in
terms the probability distribution $\Pi(\bxi,t)$, describing the
statistics of the stochastic process $\bxi(t)$. Since $\bxi(t)$ is a
linear transformation of $\bn$ we have that $\Pi(\bxi,t)\propto P_
\bn(t)$. Hence we may directly substitute $\Pi(\bxi,t)$ into the master 
equation (\ref{eq:masterapp}). We firstly note that the derivative with respect
to time in (\ref{eq:masterapp}) is taken at constant $\mathbf{n}$ so that we 
have $\mathrm{d}{\bxi}/\mathrm{d} t=-N^{1/2}\mathrm{d}\bx/\mathrm{d} t$.
This leads to
\BE\label{eq:pi}
\frac{\mathrm{d}\Pi(\bxi,t)}{\mathrm{d}t}&=&\frac{\partial
\Pi(\bxi,t)}{\partial
t}-N^{1/2}\frac{\partial\Pi(\bxi,t)}{\partial
\xi_1}\frac{\mathrm{d} x_1(t)}{\mathrm{d} t}\nonumber\\
&&-N^{1/2}\frac{\partial\Pi(\bxi,t)}{\mathrm{d}
\xi_2}\frac{\mathrm{d} x_2(t)}{\mathrm{d} t}.
\EE
Next, we write the right-hand side of Eq. (\ref{eq:masterapp}) in
terms of $\Pi(\bxi,t)$ and find that
\BE\label{eq:masterapp2}
\frac{\mathrm{d}\Pi(\bxi,t)}{\mathrm{d}t}&=&\sum_{\nu}
\left[\exp(-N^{-1/2}\mathbf{v}_{\nu}\cdot
\bm{\nabla}_{\bxi})-1\right]\nonumber\\
&& \hspace{-2em}\times
\left[Na_{\nu}(\bx(t)+N^{-1/2}\bm{\kappa})\Pi(\bxi,t)\right],
\EE
where we have also used the re-scaled reaction rates $a_{\nu}$ as introduced 
earlier, i.e.
\be\label{eq:thing}
T_{\nu}(N\bx(t)+\sqrt{N}\bm{\xi})= Na_{\nu}
\left(\bx(t)+N^{-1/2}\bm{\xi}\right).
\ee
We have exploited the continuous nature of $\bxi$ by using the shift operator:
the exponential in the differential operator,
$\bm{\nabla}_{\bxi}=\left(\frac{\partial}{\partial \xi_1},
\frac{\partial}{\partial \xi_2}\right)$, used in Eq. (\ref{eq:masterapp2}), 
has the effect of shifting the argument of the subsequent functions by the 
vector $-N^{-1/2}\bv_{\nu}$. Explicitly, for any smooth function, $F(\bxi)$,
\be
\exp(-N^{-1/2}\mathbf{v}_{\nu}\cdot
\bm{\nabla}_{\bxi})F(\bxi)=F(\bxi-N^{-1/2}\mathbf{v}_{\nu}).
\ee
Both the exponential and the rate functions, $a_{\nu}(\bx)$, in 
Eq. (\ref{eq:masterapp2}) may be expanded as polynomials in $N^{-1/2}$. We 
then take the formal limit of $N\to\infty$ keeping only terms of the two 
highest orders in $N$. One may then equate coefficients of powers of $N$ 
between the left and right-hand sides of the master equation (given by 
(\ref{eq:pi}) and (\ref{eq:masterapp2}) respectively).

To leading order one consistently recovers the mean-field
equations (\ref{eq:meanfield}), i.e.
\begin{align}
\frac{\mathrm{d}}{\mathrm{d}t}\bx(t)=\mathbf{A}(\bx(t)),
\end{align}
with $\mathbf{A}(\bx)=\sum_{\nu} \mathbf{v}_{\nu} a_{\nu}(\bx)$.
Expanding to next order one finds
\begin{align}
\frac{\partial}{\partial t}\Pi(\bm{\xi},t) =&
-\sum_{i,j}K_{ij}(t)\frac{\partial}{\partial
\xi_i}\left(\xi_j\Pi(\bm{\xi},t)\right)\nonumber\\
&+\sum_{i,j}D_{ij}(t)\frac{\partial}{\partial
\xi_i}\frac{\partial}{\partial
\xi_j}\Pi(\bm{\xi},t),\label{eq:FP}\end{align}
which is a linear Fokker-Planck equation for the distribution $\Pi(\bxi,t)$. 
The elements of the drift matrix $K(t)$ are given by 
$K_{ij}(t)=\frac{\partial}{\partial x_j}A_i(\bx(t))$ and those in the diffusion
matrix $D(t)$ are 
$D_{ij}(t)= \frac{1}{2}\sum_{\nu}(\mathbf{v}_{\nu})_{i}(\mathbf{v}_{\nu})_{j}a_{\nu}(\bx(t))$.

For the example of the Brusselator that we use, these matrices are explicitly 
given by,
\be\label{eq:Kform}
K(\bx)=\left(\begin{array}{cc}
-1-b+2cx_1 x_2 &cx_1^2\\
b-2cx_1 x_2&-cx_1^2
\end{array}\right)
\ee
and
\be
D(\bx)=\frac{1}{2}\left(\begin{array}{cc} 
1+x_1(1+b+cx_1x_2) & -x_1(b+cx_1x_2) \\ 
-x_1(b+cx_1x_2) & x_1(b+cx_1x_2) \end{array}\right) \\ 
\label{eq:Dform}
\ee

Eq. (\ref{eq:FP}) may be considered directly as a stochastic process $\bxi(t)$ 
as described by a Langevin equation. This is discussed in the main text for 
two cases. In the case of a globally stable fixed point, $\bx = \bx^{*}$,
$K(\bx)$ and $D(\bx)$ become $K^*$ and $D^*$. In the case of a limit 
cycle, $\bx = \olx(t)$, the drift and diffusion matrices are functions of time,
$K(t)$ and $D(t)$, which naturally pick up the periodicity of this solution.

\end{document}